\documentclass[prd,preprint,aps,amsfonts,eqsecnum,nofootinbib]{revtex4}
\usepackage{graphicx}
\usepackage{amsmath}
\begin {document}

\preprint{WM-03-104}

\title{Phenomenology of Lorentz-Conserving Noncommutative QED}
\author{Justin M. Conroy}
 \email{jconroy@camelot.physics.wm.edu}
\author{Herry J. Kwee}%
 \email{herry@camelot.physics.wm.edu}
\author{Vahagn Nazaryan}
 \email{vrnaza@wm.edu}
\affiliation{%
Nuclear and Particle Theory Group, Department of Physics\\
College of William and Mary, Williamsburg, VA 23187-8795}%

\date{\today}

\begin{abstract}
Recently a version of Lorentz-conserving noncommutative field theory
(NCFT) has been suggested. The underlying Lie algebra of the theory is the
same as that of Doplicher, Fredenhagen, and Roberts.  In
Lorentz-conserving NCFT the matrix parameter ${\theta}^{\mu \nu}$ which
characterizes the canonical NCFT's is promoted to an operator $ \hat
{\theta}^{\mu \nu}$ that transforms as a Lorentz tensor.  In this paper, we
calculate phenomenological consequences of the QED version of this theory
by looking at various collider processes.  In particular we calculate
modifications to M\o ller scattering, Bhabha scattering, $e^+e^-
\rightarrow \mu^+ \mu^-$ and $ e^+e^- \to \gamma \gamma $.  We obtain
bounds on the noncommutativity scale from the existing experiments at LEP
and make predictions for what may be seen in future collider experiments.
\end{abstract}

\maketitle
\section{\label{sec:intro}Introduction}

It is interesting to consider the possibility that the structure of
space-time is nontrivial.  In one of the most popular scenarios position
four-vectors are promoted to
operators that do not commute at short distance scales~\cite{Madore,HAY,
ARM,LIA,CSJT,Jurco,Calmet,lvncsm,MPR,Chaichian,HK,CHK,CCL,ABDG,Carlson:2002zb,
VN,h2,HPR,mncqed,ncxd,Iltan,CCZ,Morita,dop,snyder}.
There has been a lot of work on field theories with an underlying
noncommutative space-time structure. Jur\v{c}o $et$ $al.$~\cite{Jurco} have
presented a formalism on how to construct non-Abelian gauge theories in
noncommutative spaces from a consistency relation. Using a similar
approach Carlson, Carone and Zobin (CCZ)~\cite{CCZ} have formulated 
noncommutative Lorentz-conserving QED based on a contracted Snyder~\cite{snyder}
algebra, thus offering a general prescription as how to formulate
noncommutative Lorentz-conserving gauge theories.  In this algebra the
selfadjoint spacetime coordinate operators satisfy the following 
commutation relation,
\begin{equation}
[\hat x^{\mu},\hat x^{\nu}] = i\hat {\theta}^{\mu \nu}.
\label{DFR}
\end{equation}
Here ${\hat \theta}^{\mu \nu} = - {\hat \theta}^{\nu \mu} $ transforms as
a Lorentz tensor and is in the same algebra with $\hat x^{\mu}$. This
algebra is Lorentz covariant.

The Lie algebra considered by CCZ is the same as the Lie algebra of
Doplicher, Fredenhagen, and Roberts (DFR)~\cite{dop}.  Interestingly
enough DFR came to the formulation of their algebra by considering
modifications of spacetime structure in theories that are designed
to quantize gravity.  The DFR algebra places limitations on the precision 
of localization in spacetime. As noted in~\cite{dop}, quantum spacetime
can be regarded as a novel underlying geometry for a quantum field theory
of gravity.

Interest in noncommutative spacetime originated with the work of 
Connes and collaborators~\cite{Connes} and has gained more attention due to 
developments in string theory~\cite{SW}, where noncommutative spacetime has 
been shown to arise in a low energy limit.
In string theories ${\theta}^{\mu \nu}$ is just an antisymmetric c-number.
Theories involving noncommutative spacetime structure based on algebras
with c-number ${\theta}^{\mu \nu}$ suffer from Lorentz-violating effects.
Such effects are severely
constrained~\cite{MPR,Chaichian,HK,CHK,CCL,ABDG,Carlson:2002zb,VN,h2}
by a variety of low energy experiments~\cite{LeExp}.
Lorentz-violating effects appear in field theories as a consequence of ${\theta}^{0i}$
and ${\epsilon}^{ijk}{\theta}^{ij}$ defining preferred direction in a
given Lorentz frame.  In contrast to this the noncommutative QED (NCQED)
formulated by CCZ based on Eq.~(\ref{DFR}) is free from Lorentz-violating
effects.
     
Carlson, Carone and Zobin have connected the DFR Lie algebra Eq.~(\ref{DFR}),
and the antisymmetric tensor $\hat {\theta}^{\mu \nu}$ to experimental
observables, by showing how to formulate a quantum field theory on this 
noncommutative spacetime. Similar issues have been discussed by Morita 
$et$ $al.$~\cite{Morita}. These theories make it possible to study phenomenological 
consequences of Lorentz-conserving noncommutative spacetime. As a beginning, 
CCZ have studied light-by-light elastic scattering and obtained
contributions that can be significant with respect to the standard model background.

In this paper we calculate other phenomenological consequences of 
Lorentz-conserving NCQED formulated by CCZ. We consider various
collider processes such as Bhabha and M\o ller scattering, $e^+e^- \to
\mu^+ \mu^-$ and $ e^+e^- \to \gamma \gamma $.  The experiments at 
planned colliders will provide means of testing the properties and the
structure of space-time at smaller distance scales.  We note that any
property prescribed to space-time, if confirmed experimentally, must 
affect all interactions.

In the following section we discuss the underlying formalism of
noncommutative Lorentz-conserving gauge theories, with emphasis on
NCQED.  In Section III we study the Lorentz-conserving 
NCQED by considering various collider processes. In
Section IV we obtain bounds on the noncommutativity scale 
from Bhabha scattering, $e^+e^- \to \mu^+ \mu^-$
and $e^+e^- \to \gamma \gamma$ experiments.  We summarize our discussion
in Section V with some concluding remarks.

\section{\label{sec:Theory}Algebra and QED Formulation}

The simplest construction of a Lorentz-conserving noncommutative
theory involves promoting the position four-vector to an operator
which satisfies the DFR Lie algebra

\begin{eqnarray}
& [\hat{x}^{\mu},\hat{x}^{\nu}] &= i\hat{\theta}^{\mu\nu} ,\nonumber  \\ 
& [\hat{\theta}^{\mu\nu},\hat{x}^{\lambda}] &= 0 ,  \nonumber \\       
& [\hat{\theta}^{\mu\nu},\hat{\theta}^{\alpha\beta}] &= 0 ,  \label{eq:DFR}
\end{eqnarray}
where $\theta^{\mu\nu}$ is antisymmetric and transforms as a Lorentz
tensor.

On the other hand, CCZ took as the starting point Snyder's algebra,
\begin{eqnarray}
& [\hat{x}^{\mu},\hat{x}^{\nu}]=i a^{2}\hat{M}^{\mu\nu},\nonumber \\ &
[\hat{M}^{\mu\nu},\hat{x}^{\lambda}]=i(\hat{x}^{\mu}g^{\nu\lambda}-\hat{x}^{\nu}g^{\mu\lambda}),
\nonumber \\ &
[\hat{M}^{\mu\nu},\hat{M}^{\alpha\beta}]=i(\hat{M}^{\mu\beta}g^{\nu\alpha}
+
\hat{M}^{\nu\alpha}g^{\mu\beta}-\hat{M}^{\mu\alpha}g^{\nu\beta}-\hat{M}^{\nu\beta}g^{\mu\alpha})
\label{eq:snyder} .
\end{eqnarray}
 Snyder's algebra (which is the same as the algebra of SO(4,1))
describes a Lorentz-invariant noncommutative discrete spacetime
characterized by a fundamental length scale $a$.  By constructing an
explicit representation for $\hat{x}$ and $\hat{M}$ in terms of
differential operators, the Lorentz invariance of
Eq.~(\ref{eq:snyder}) was demonstrated \cite{snyder}.  CCZ then
extracted the DFR Lie algebra by performing a particular contraction
on Eq.~(\ref{eq:snyder}).  Specifically, by rescaling
$M^{\mu\nu}=\hat{\theta}^{\mu\nu}/b$ and holding the ratio $a^{2}/b=1$
fixed, the limit $b \rightarrow 0$, $a \rightarrow 0$ yields the DFR
Lie algebra.  Thus, the Lorentz covariance of Snyder's Lie algebra
implies the Lorentz covariance of Eq.~(\ref{eq:DFR}) \cite{CCZ}.  The
commutator of $\hat{\theta}^{\mu\nu}$ and $\hat{M}^{\mu\nu}$ is
\begin{equation}
[\hat{M}^{\mu\nu},\hat{\theta}^{\alpha\beta}]=i(\hat{\theta}^{\mu\beta}g^{\nu\alpha}
+
\hat{\theta}^{\nu\alpha}g^{\mu\beta}-\hat{\theta}^{\mu\alpha}g^{\nu\beta}-\hat{\theta}^{\nu\beta}g^{\mu\alpha}) ,
\end{equation}
as one would expect if $\hat{\theta}^{\mu\nu}$ is a Lorentz tensor.
Note that the contraction also implies that the eigenvalues of the
position operator of the DFR algebra are continuous.

To develop a field theory on a noncommutative spacetime, one defines a
one-to-one mapping which associates functions of the noncommuting
coordinates with functions of the typical c-number coordinates.  In
the canonical noncommutative theory this is achieved via a Fourier
transform

\begin{equation}
\hat{f}(\hat{x})=\frac{1}{{2\pi}^{n}}\int d^{n}k\, e^{-ik\hat{x}} \int
d^{n}x\, e^{ikx} f(x). \label{eq:fourier}
\end{equation}

In the Lorentz-conserving case the presence of the operator
$\hat{\theta}^{\mu\nu}$ requires that the mapping involve a new
c-number coordinate $\theta^{\mu\nu}$ (no hat).  Functions of the
noncommuting coordinates are then related to functions of c-number
coordinates by
\begin{equation}
\hat{f}(\hat{x},\hat{\theta})=\int \frac{d^{4}\alpha}{(2
\pi)^{4}}\frac{d^{6}B}{(2 \pi)^{6}}e^{-i(\alpha_{\mu}
\hat{x}^{\mu}+\frac{B_{\mu\nu}\hat{\theta}^{\mu\nu}}{2})}
\tilde{f}(\alpha, B) \label{eq:four1} ,
\end{equation}
where
\begin{equation}
\tilde{f}(\alpha, B)=\int d^{4}x d^{6}\theta e^{i(\alpha_{\mu}
x^{\mu}+\frac{B_{\mu\nu}\theta^{\mu\nu}}{2})}f(x,
\theta). \label{eq:four2}
\end{equation}
Lorentz invariance requires that $B$ transform as a two index Lorentz tensor.

To ensure that operator multiplication be preserved,
$\hat{f}\hat{g}=\widehat{f\star g}$, one finds that the rule for
ordinary multiplication must be modified:
\begin{equation}
(f\star g)(x, \theta)=f(x, \theta)
\exp[\frac{i}{2}\stackrel{\leftarrow}{\partial}_{\mu}
\theta^{\mu\nu}\stackrel{\rightarrow}{\partial_{\nu}}]g(x,\theta).
\end{equation}
The $\theta$ dependence of the functions distinguishes this result
from the $\star$-product of the canonical noncommutative theory.
Eqs.~(\ref{eq:four1}) and (\ref{eq:four2}) allow one to work solely
with functions of classical coordinates $x$ and $\theta$, provided
that all multiplication be promoted to a $\star$-product.

The introduction of a Lorentz invariant weighting function $W(\theta)$
allows for the following generalization of the operator trace:
\begin{equation}
 {\rm Tr} \hat{f}=\int d^{4}x\, d^{6}\theta \, W(\theta)f(x, \theta).  \label{eq:Tr}
\end{equation}
In \cite{CCZ} CCZ took the normalization to be
\begin{equation}
\int d^{6}\theta\, W(\theta)=1 .
\end{equation}
It is straightforward to demonstrate the cyclic property of
Eq.~(\ref{eq:Tr}), \textit{i.e.} ${\rm Tr} \hat{f} \hat{g}={\rm Tr} \hat{g} \hat{f}$.
One requires that for large $|\theta^{\mu\nu}|$, $W(\theta)$ dies off
sufficiently fast in order that all integrals be well defined
\cite{CCZ}.  Lorentz-invariance requires that $W$ be an even function
of $\theta$, which yields
\begin{equation}
\int d^{6}\theta\, W(\theta)\,\theta^{\mu\nu}=0.  \label{eq:even}
\end{equation}
As will be seen, this restriction has interesting consequences on
possible collider signatures of the theory.

Field theory interactions are extracted by performing the
$d^{6}\theta$ integral, resulting in the action
\begin{equation}
\mathcal{S}=Tr \hat{\mathcal{L}}= \int d^{4}x\, d^{6} \theta\, W(\theta)\,
\mathcal{L}(\phi, \partial \phi)_{\star} \, ,
\end{equation}
where the notation in $\mathcal{L}(\phi, \partial \phi)_{\star}$
indicates $\star$-product multiplication.

As was mentioned, in the Lorentz-conserving noncommutative theory
the initial ``fields'' are generally functions of $x$ and $\theta$,
and must be related to ordinary quantum fields which are only
functions of $x$ .  CCZ showed how this can be done for NCQED using a
nonlinear field redefinition and an expansion in $\theta$.  Since the
phenomenology of NCQED is the topic of this paper, all developments will
be directed toward a U(1) gauge theory.  For completeness the formalism
presented in \cite{CCZ} is reviewed.

In Lorentz-conserving NCQED, one has a matter field $\psi$ and gauge field
$A$ .  For a U(1) gauge transformation characterized by a parameter
$\Lambda(x, \theta)$, the fields transform as
\begin{equation}
\psi(x, \theta) \rightarrow U \star \psi(x, \theta),
\end{equation}
and
\begin{equation}
A_{\mu}(x, \theta) \rightarrow U \star A_{\mu}(x, \theta) \star
U^{-1}+\frac{i}{e} U \star \partial_{\mu}U^{-1} ,
\end{equation}
where

\begin{eqnarray}
\lefteqn{U =(e^{i\Lambda})_{\star}} \nonumber \\ & & =1+i
 \Lambda(x,\theta)+\frac{1}{2!} i\Lambda(x,\theta) \star
 i\Lambda(x,\theta)+...      .
\end{eqnarray}
A U(1) gauge invariant Lagrangian is
\begin{equation}
\mathcal{L}=\int d^{6} \theta W(\theta)[-\frac{1}{4}F_{\mu\nu} \star
F^{\mu\nu}+\bar{\psi} \star (i \not\!\!{D}-m)\star \psi], \label{eq:QED1}
\end{equation}
where 
\begin{equation} 
D_{\mu}=\partial_{\mu}-ie A_{\mu}, \label{eq:QED2}
\end{equation}
and the field strength is
\begin{equation}
F_{\mu\nu}=\partial_{\mu}A_{\nu}-\partial_{\nu}A_{\mu}-ie
[A_{\mu}\stackrel{\star}{,}A_{\nu}] . \label{eq:QED3}
\end{equation}
In demonstrating the gauge invariance of Eq.~(\ref{eq:QED1}) and the
cyclic property of Eq.~(\ref{eq:Tr}), the following identity is useful
\begin{equation}
\int d^{4}x f \star g=\int d^{4}x f g .
\end{equation}
Eqs. (\ref{eq:QED1}),~(\ref{eq:QED2}), and~(\ref{eq:QED3}) are similar in
form to those obtained in the canonical NCQED case, the difference
again being the $\theta$ dependence of the fields $\psi(x,\theta)$ and
$A(x, \theta)$ in Eq.~(\ref{eq:QED1}).  One must have a way of
relating $\psi$ and $A$ to ordinary quantum fields which are only
functions of $x$.  This is accomplished by utilizing the behavior of
the weighting function Eq.~(\ref{eq:Tr}), which allows an expansion of the
fields and gauge parameter in powers of $\theta$.  A similar technique
involving field expansions was first used in constructing a
noncommutative SU($N$) gauge theory in \cite{Jurco}.  The coefficients
of the power series are thus only functions of $x$ and correspond to
ordinary quantum fields.  From requirements of gauge invariance and
noncommutativity, these coefficients can be determined order by order
in $\theta$.

The matter field, gauge field, and gauge parameter of NCQED are expanded
as:
\begin{equation}
\Lambda_{\alpha}(x,\theta) = \alpha(x) + \theta^{\mu\nu}
\Lambda_{\mu\nu}^{(1)} (x;\alpha) + \theta^{\mu\nu} \theta^{\eta\sigma}
\Lambda_{\mu\nu\eta\sigma}^{(2)} (x;\alpha) + \cdots,
\end{equation}
\begin{equation}
A_{\rho}(x,\theta) = A_{\rho}(x) + \theta^{\mu\nu} A_{\mu\nu\rho}^{(1)}(x)
+ \theta^{\mu\nu} \theta^{\eta\sigma} A_{\mu\nu\eta\sigma\rho}^{(2)}(x) +
\cdots,
\end{equation}
\begin{equation}
\psi(x,\theta) = \psi(x) + \theta^{\mu\nu} \psi_{\mu\nu}^{(1)} +
\theta^{\mu\nu} \theta^{\eta\sigma} \psi_{\mu\nu\eta\sigma}^{(2)} (x) +
\cdots.
\end{equation}
The lowest order term in each expansion corresponds to the ordinary
QED term.  Thus, ordinary QED can be extracted by taking the
commutative limit, $\theta^{\mu\nu} \rightarrow 0$.

Consider an infinitesimal transformation of a matter field $\psi(x)$
in an ordinary U(1) gauge theory:
\begin{equation}
\delta_{\alpha} \psi(x)=i \alpha(x)\psi(x) \label{eq:matter} .
\end{equation}
For a Lorentz-conserving noncommutative theory, this is generalized to
\begin{equation}
\delta_{\alpha}\psi(x,\theta)=i \Lambda_{\alpha}(x,\theta)\star
\psi(x,\theta) \label{eq:trans3} .
\end{equation}
In an Abelian gauge theory two successive gauge transformations must
then satisfy the relation
\begin{equation}
(\delta_{\alpha}\delta_{\beta}-\delta_{\beta}\delta_{\alpha})\psi(x,\theta)=0
.
\label{eq:trans1} 
\end{equation}

For Eq.(\ref{eq:trans1}) to hold, $\Lambda$ must satisfy
\begin{equation}
i\delta_{\alpha}\Lambda_{\beta}-i\delta_{\beta}\Lambda_{\alpha}+[\Lambda_{\alpha}\stackrel{\star}{,}\Lambda_{\beta}]=0
\label{eq:trans2}.
\end{equation}
The parameter $\Lambda$ can then be determined at each order in
$\theta$.  Specifically, it can be shown that
\begin{equation}
\Lambda_{\mu\nu}^{(1)}(x;\alpha)=\frac{e}{2}\partial_{\mu}\alpha(x)A_{\nu}(x)
\label{eq:T1}
\end{equation}
and
\begin{equation}
\Lambda_{\mu\nu\eta\sigma}^{(2)}(x;\alpha)=-\frac{e^{2}}{2}\partial_{\mu}\alpha(x)A_{\eta}(x)\partial_{\sigma}A_{\nu}(x)
\label{eq:T2}
\end{equation}
satisfy the condition of Eq.~(\ref{eq:trans2}).  The gauge and matter
fields are treated in a similar manner.

The restriction of a gauge field transforming infinitesimally as
\begin{equation}
\delta_{\alpha}A_{\sigma}=\partial_{\sigma}\Lambda_{\alpha}+i[\Lambda_{\alpha}\stackrel{\star}{,}A_{\sigma}]
,
\end{equation}
is satisfied by the following expressions for $A^{(1)}$ and $A^{(2)}$:
\begin{equation}
A_{\mu\nu\rho}^{(1)}(x)=-\frac{e}{2}A_{\mu}(\partial_{\nu}A_{\rho}+F_{\nu\rho}^{0})
, \label{eq:T3}
\end{equation}
\begin{equation}
A_{\mu\nu\eta\sigma\rho}^{(2)}(x)=\frac{e^2}{2}(A_{\mu}A_{\eta}\partial_{\sigma}F_{\nu\rho}^{0}-\partial_{\nu}A_{\rho}\partial_{\eta}A_{\mu}A_{\sigma}+A_{\mu}F_{\nu\eta}^{0}F_{\sigma\rho}^{0})
\label{eq:T4} ,
\end{equation}
where
\begin{equation}
F_{\mu\nu}^{0}=\partial_{\mu}A_{\nu}-\partial_{\nu}A_{\mu}
\end{equation}
 is the ordinary QED field strength tensor.

Likewise, one can show that for a matter field transforming
infinitesimally as Eq.~(\ref{eq:trans3}), the appropriate forms of
$\psi^{(1)}$ and $\psi^{(2)}$ are
\begin{equation}
\psi_{\mu\nu}^{(1)}(x)=-\frac{e}{2}A_{\mu}\partial_{\nu}\psi  \label{eq:T5}
\end{equation}
and
\begin{eqnarray}
\psi_{\mu\nu\eta\sigma}^{(2)}(x)=\frac{e}{8}(-i\partial_{\mu}A_{\eta}\partial_{\nu}\partial_{\sigma}\psi+e
A_{\mu}A_{\eta}\partial_{\nu}\partial_{\sigma}\psi+2e
A_{\mu}\partial_{\nu}A_{\eta}\partial_{\sigma}\psi \nonumber \\ +e
A_{\mu}F_{\nu\eta}^{0}\partial_{\sigma}\psi-\frac{e}{2}\partial_{\mu}A_{\eta}\partial_{\nu}A_{\sigma}\psi+i
e^{2} A_{\mu}A_{\sigma}\partial_{\eta}A_{\nu} \psi) . \label{eq:T6}
\end{eqnarray}

Interactions are extracted by substituting Eqs. (\ref{eq:T1}),~
(\ref{eq:T2}),~(\ref{eq:T3}),~(\ref{eq:T4}),~(\ref{eq:T5}),~(\ref{eq:T6})
into the Lagrangian Eq.~(\ref{eq:QED1}).  We expand the Lagrangian through
$\theta^{2}$ and evaluate the $d^{6}\theta$ integral using the
weighted average
\begin{equation}
\int d^{6}\theta W(\theta) \theta^{\mu\nu}
\theta^{\eta\rho}=\frac{\langle\theta^{2}\rangle}{12}(g^{\mu\eta}g^{\nu\rho}-g^{\mu\rho}g^{\eta\nu})
,
\end{equation}
where the expectation value is defined as
\begin{equation}
\langle \theta^{2}\rangle \equiv \int d^{6}\theta W(\theta) \theta_{\mu\nu}\theta^{\mu\nu} .
\end{equation}
It is natural to define $\Lambda_{NC} = (12/\langle \theta^{2} \rangle)^{1/4}$
 which characterizes the energy scale where noncommutative effects become
 relevant.  The restriction on $W$ from Eq.~(\ref{eq:even}) demands that
 only terms containing even powers of $\theta$ will result in
 interaction vertices.  Thus, for example, the three-photon vertex of
 canonical NCQED is not present.  The next section focuses on the
 phenomenology of a U(1) theory whose spacetime coordinate operators
 obey the DFR Lie algebra.  Possible collider signatures are considered
 and bounds on the energy scale $\Lambda_{NC}$ are obtained.

\section{\label{sec:NCQED}Collider Signatures}

The Lagrangian for QED with Lorentz-invariant noncommutative spacetime
Eq.~(\ref{eq:QED1}) can be written as an expansion in $\theta$ order by order
using the nonlinear field redefinition described above.  The zeroth order
in $\theta$ will give the ordinary QED Lagrangian.  The first order is zero
due to the evenness of the weighting function $W(\theta)$.  The first
nontrivial contributions come from the second order, they include:
\begin{enumerate}
\item the 4-photon vertex, which has been discussed extensively in \cite{CCZ},
\item the correction to 2-fermion-1-photon vertex (ordinary QED vertex),
\item the 2-fermion-2-photon vertex.
\end{enumerate}

The lowest order correction to the ordinary QED vertex comes from the
following terms in Lagrangian density:
\begin{eqnarray}
\bar\psi^{(2)} (i\not\!\partial-m)\psi^{(0)} + \bar\psi^{(0)}
(i\not\!\partial-m)\psi^{(2)} \nonumber \\
+ \frac{e}{2} \{ (\bar\psi^{(0)}\star\not\!\! A^{(0)}) \psi^{(0)} + \bar\psi^{(0)}
(\not\!\!A^{(0)}\star\psi^{(0)}) \}, \label{eq:vertex3}
\end{eqnarray}\\
where we retain only the second order term in contributions to the
$\star$-product shown in the last two terms.  The first two terms will
go to zero if both fermion fields are on shell.  And the
2-fermion-2-photon vertex comes from:
\begin{widetext}
\begin{eqnarray}
\bar\psi^{(2)} (i\not\!\partial-m)\psi^{(0)} &+& \bar\psi^{(0)}
(i\not\!\partial-m)\psi^{(2)} + \bar\psi^{(1)}
(i\not\!\partial-m)\psi^{(1)}\nonumber\\
+ \quad e \{ \bar\psi^{(2)}\not\!\!A^{(0)}
\psi^{(0)} &+& \bar\psi^{(0)} \not\!\!A^{(0)}\psi^{(2)} \} \nonumber \\
+ \quad e \{ (\bar\psi^{(0)}\star\not\!\!A^{(0)}) \psi^{(1)} &+& \bar\psi^{(1)}
(\not\!\!A^{(0)}\star\psi^{(0)}) + (\bar\psi^{(0)}\star\not\!\!A^{(1)})
\psi^{(0)} \}, \label{eq:vertex4}
\end{eqnarray}
\end{widetext}
where this time we retain only the first order in the $\star$-product
shown.

\subsection{\label{sec:dilepton}Dilepton Production, $e^+e^-\to l^+l^-$}

First we consider processes in which all fermions are on shell, {\it i.e.}
dilepton production $e^+e^-\to l^+l^-$.  For processes up to tree level
Feynman diagram, only
\begin{equation}
\frac{e}{2} \{ (\bar\psi^{(0)}\star\not\!\!A^{(0)}) \psi^{(0)} + \bar\psi^{(0)}
(\not\!\!A^{(0)}\star\psi^{(0)}) \} \nonumber
\end{equation}
will contribute to the vertex correction since all the fermions are on
shell.  This Lagrangian term reduces to:
\begin{eqnarray}
\frac{e}{2} \frac {\langle\theta^2\rangle}{96}
\{\bar\psi (\partial_\mu \partial_\nu\not\!\!A)(\partial^\mu
\partial^\nu \psi) + (\partial^\mu \partial^\nu\bar\psi) (\partial_\mu
\partial_\nu\not\!\!A) \psi \}.
\end{eqnarray}
\begin{figure}
\scalebox{.4}{\includegraphics{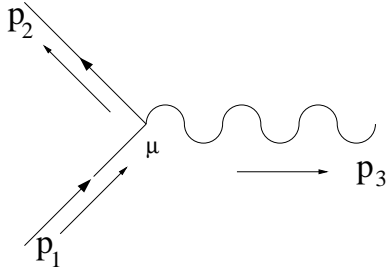}}
\caption{\label{fig:vertex3}2-fermions-1-photon vertex.}
\end{figure}
\\
From this we obtain the following Feynman rule for the 2-fermion-1-photon
vertex with all fermions on shell and with momenta labeled as in
Fig.~\ref{fig:vertex3}:
\begin{eqnarray}
ie \{1+\frac{\langle\theta^2\rangle}{384} (p_3)^4\} \gamma^\mu,
\label{eq:feynrule1}
\end{eqnarray}
where we have not made the assumption that the fermions are
massless (although we do set $m=0$ in the cross section formula).

We will consider the following processes which are affected by this vertex
correction: Bhabha scattering, $e^+e^-\to\mu^+\mu^-$ and M\o ller scattering.
\begin{figure*}
\scalebox{.3}{\includegraphics{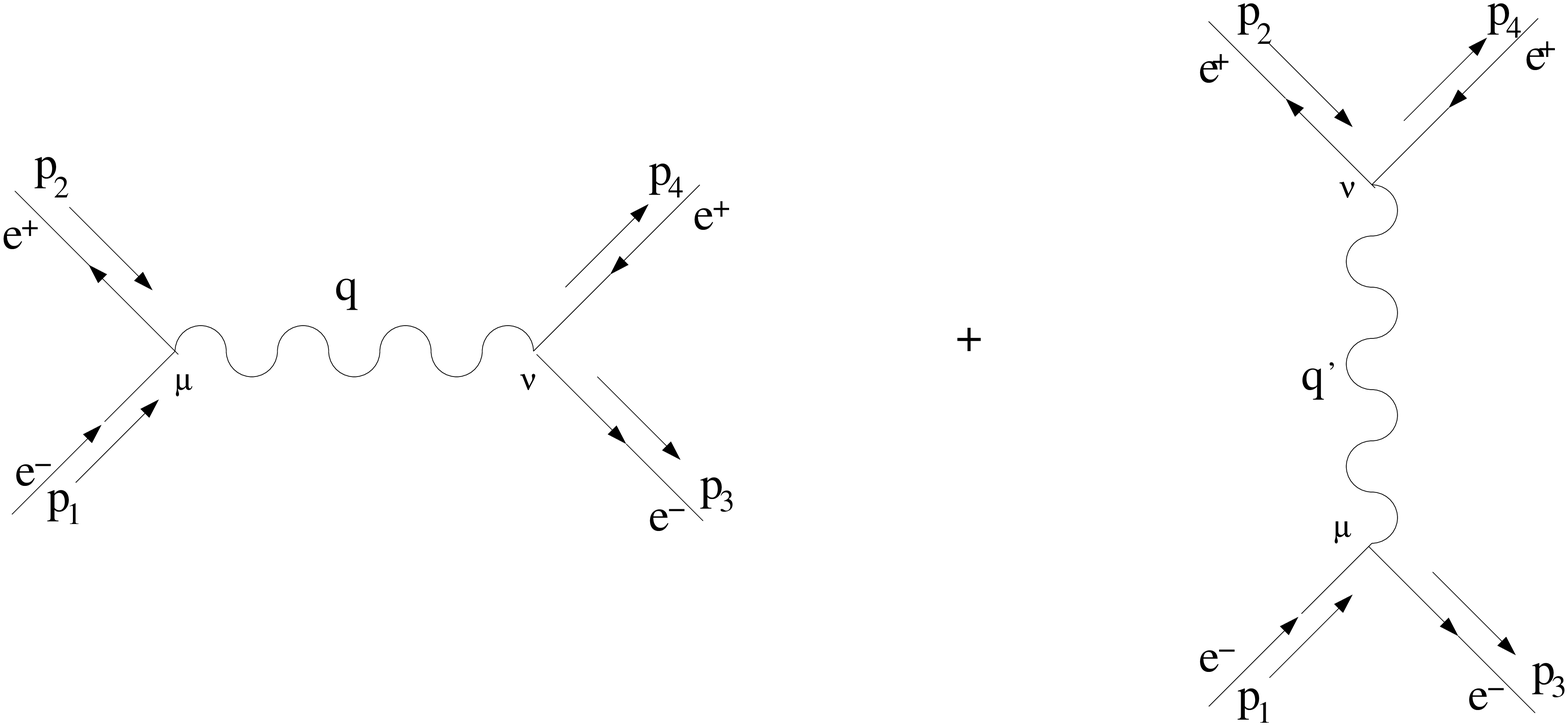}}
\caption{\label{fig:bhabha}Bhabha Scattering.}
\end{figure*}
The matrix element with vertex correction for Bhabha scattering
(Fig.~\ref{fig:bhabha}) is:
\begin{eqnarray}
i \mathcal{M} = \bar u(p_3)(ie\gamma^\nu)
(1&+&\frac{\langle\theta^2\rangle}{384}q^4) v(p_4)
\frac{-ig_{\mu\nu}}{q^2+i\epsilon} \nonumber \\
\times\bar v(p_2) (ie\gamma^\mu)
(1&+&\frac{\langle\theta^2\rangle}{384}q^4) u(p_1) \nonumber \\
- \bar v(p_2)(ie\gamma^\nu)
(1&+&\frac{\langle\theta^2\rangle}{384}q'^4) v(p_4)
\frac{-ig_{\mu\nu}}{q'^2+i\epsilon} \nonumber \\
\times\bar u(p_3) (ie\gamma^\mu)
(1&+&\frac{\langle\theta^2\rangle}{384}q'^4) u(p_1).
\end{eqnarray}\\
Squaring the matrix element and summing(averaging) over the final(initial)
fermion spin states will give:
\begin{equation}
\overline{|\mathcal{M}|^{\rm2}} = 2e^4 \{F_s^2 (\frac{t^2 + u^2}{s^2}) +
2F_sF_t\frac{u^2}{st} + F_t^2(\frac{u^2 + s^2}{t^2})\}, \label{eq:msqr1}
\end{equation}\\
where we define $F_s = \{1+\frac{\langle\theta^2\rangle}{96} {\frac{s^2}{4}} \}^2$
with s,t and u are the Mandelstam variables.  To first order in
$\langle\theta^2\rangle/12$ this will give us the center of mass
(CM) differential cross section:
\begin{equation}
\frac{d\sigma}{d\cos\theta} = \left(
\frac{d\sigma}{d\cos\theta}\right)_{QED} +
\frac{\pi\alpha^2}{s} \frac{\langle\theta^2\rangle}{96} \{s^2 + t^2 + 2u^2 +
u^2(\frac{t}{s} + \frac{s}{t}) \}, \label{eq:xsct1}
\end{equation}
where $\theta$ is the CM scattering angle.

The same results for $e^+e^-\to\mu^+\mu^-$ can be obtained easily by just
throwing away the $t$ channel in the Bhabha scattering calculation, assuming
the muons are massless.  The spin average square matrix element is:
\begin{equation}
\overline{|\mathcal{M}|^{\rm2}} = 2e^4 F_s^2 (\frac{t^2 + u^2}{s^2}). \label{eq:msqr3}
\end{equation}\\
And to first order in $\langle\theta^2\rangle/12$ this will give us:
\begin{equation}
\frac{d\sigma}{d\cos\theta} = \left(
\frac{d\sigma}{d\cos\theta}\right)_{QED} (1 + \frac{\langle\theta^2\rangle}{96} s^2). \label{eq:xsct3}
\end{equation}

\subsection{\label{sec:moller}M\o ller Scattering}

For M\o ller scattering, the spin average square matrix element is
obtained by using crossing symmetry from Bhabha scattering:
\begin{equation}
\overline{|\mathcal{M}|^{\rm2}} = 2e^4 \{F_t^2 (\frac{u^2 + s^2}{t^2}) +
2F_tF_u\frac{s^2}{tu} + F_u^2(\frac{s^2 + t^2}{u^2}) \}. \label{eq:msqr2}
\end{equation}\\
To first order in $\langle\theta^2\rangle/12$ this gives us the CM
differential cross section:
\begin{equation}
\frac{d\sigma}{d\cos\theta} = \left(
\frac{d\sigma}{d\cos\theta}\right)_{QED} +
\frac{\pi\alpha^2}{s}\frac{\langle\theta^2\rangle}{96} \{t^2 + u^2 + 2s^2 +
s^2(\frac{u}{t}  + \frac{t}{u}) \}. \label{eq:xsct2}
\end{equation}

\subsection{\label{sec:diphoton}Diphoton Production, $e^+e^-\to\gamma\gamma$}

In order to calculate the cross section for $e^+e^-\to\gamma\gamma$,
we first need to calculate the full correction to ordinary QED vertex, not
just the case when all fermions are on shell.  This requirement comes from
the fact that in diphoton production we have fermion propagators in the
Feynman diagrams.  By using the non-linear field redefinition for
$\psi^{(2)}$, the Lagrangian for the full correction can be written as:
\begin{widetext}
\begin{eqnarray}
ie \frac {\langle\theta^2\rangle}{96} \left[ (\partial_\mu A^\mu) \{(\partial^2
\bar\psi) (i\not\!\partial-m) \psi \right. &+&\left. \{(i\partial_\alpha+m)
\bar\psi \} \gamma^\alpha (\partial^2\psi)\} \nonumber \right. \\
\left. -(\partial_\mu A_\nu) \{ (\partial^\mu \partial^\nu \bar\psi)
(i\not\!\partial-m) \psi \right. &+& \left.
\{(i\partial_\alpha+m)\bar\psi\} \gamma^\alpha
(\partial^\mu \partial^\nu \psi)\} \nonumber \right. \\
\left. - \frac{i}{2} \{\bar\psi (\partial_\mu \partial_\nu\not\!\!A)
(\partial^\mu\partial^\nu\psi) \right. &+& \left. (\partial^\mu
\partial^\nu\bar\psi)(\partial_\mu\partial_\nu\not\!\!A) \psi \} \right].
\end{eqnarray}
\end{widetext}
Then the Feynman rule for the 2-fermion-1-photon vertex with all fermions and
photons possibly off-shell is (Fig.~\ref{fig:vertex3}):
\begin{eqnarray}
ie\{\gamma^\mu + \frac{\langle\theta^2\rangle}{96} \left[ (\not\!p_1-m)p_2^2 p_3^\mu
\right. &-& \left. (\not\!p_2-m) p_1^2 p_3^\mu \nonumber \right. \\
\left. + (\not\!p_2-m)(p_1.p_3)p_1^\mu \right. &-& \left.
(\not\!p_1-m)(p_2.p_3)p_2^\mu \nonumber\right. \\
\left. + \frac{1}{2} \{(p_1.p_3)^2 \right. &+& \left.
(p_2.p_3)^2\}\gamma^\mu \right]\}.
\end{eqnarray}

Next we need to calculate the contribution from the new vertex, {\it
i.e.}, 2-fermion-2-photon vertex.  The Lagrangian for this vertex is:
\begin{eqnarray}
i e^2 \frac {\langle\theta^2\rangle}{96} \left[ A_\mu (\partial_\alpha
A_\nu) \{ (\partial^\mu \bar\psi) \gamma^\alpha (\partial^\nu \psi)
\right. &-& \left. (\partial^\nu \bar\psi) \gamma^\alpha (\partial^\mu
\psi \} \nonumber \right. \\
- \left. (\partial_\mu A_\nu) \{ (\partial^\mu\partial^\nu \bar\psi)
\not\!\!A \psi \right. &-& \left. \bar\psi \not\!\!A (\partial^\mu
\partial^\nu \psi)\} \nonumber \right. \\
+ \left. 2A_\mu F_{\nu\alpha} \{ (\partial^\mu \bar\psi)\gamma^\alpha
(\partial^\nu\psi) \right. &-& \left. (\partial^\nu \bar\psi)\gamma^\alpha
(\partial^\mu \psi\} \right],
\end{eqnarray}
and we put all the fermions and photons on shell to simplify the
calculation.  This simplification is possible since in the calculation for
diphoton production up to second order in $\theta$ for the
2-fermion-2-photon vertex all fermions and photons are on shell.
\begin{figure}
\scalebox{.3}{\includegraphics{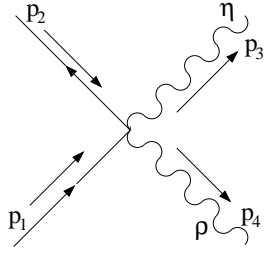}}
\caption{\label{fig:vertex4}Two fermions - two photon vertex.}
\end{figure}
Labeling momenta as in Fig.~\ref{fig:vertex4}, we obtain the Feynman rule
for the 2-fermion-2-photon vertex with all fermions and photons on shell:
\begin{eqnarray}
{ie^2} \frac{\langle\theta^2\rangle}{96} \left[(p_1.p_3)\{p_2^\rho
\gamma^\eta \right. &-& \left. p_1^\eta \gamma^\rho\} \nonumber \right. \\
\left. + (p_1.p_4)\{p_2^\eta \gamma^\rho \right. &-& \left. p_1^\rho
\gamma^\eta\} \nonumber \right. \\
\left. + (\not\!p_3-\not\!p_4) \{p_1^\rho p_2^\eta \right. &-& \left.
p_1^\eta p_2^\rho\} \right].
\end{eqnarray}

\begin{figure*}
\scalebox{.25}{\includegraphics{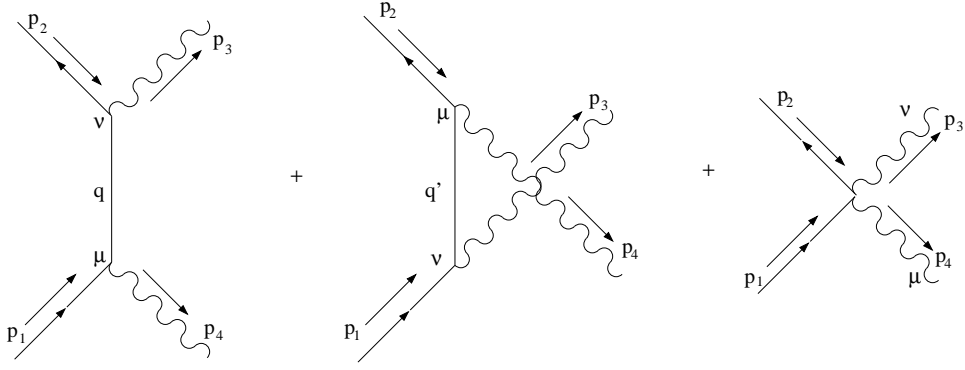}}
\caption{\label{fig:diphoton}Feynman diagrams for $e^+e^-\to\gamma\gamma$}
\end{figure*}

Putting all these rules together, the cross section up to first order in
$\langle\theta^2\rangle/12$ for diphoton production can be calculated
(Fig.~\ref{fig:diphoton}).
The matrix element for diphoton production can be written as the sum of
the three diagrams: $i \mathcal{M} = \mathit{i} \mathcal{M}_{\rm1} +
\mathit{i}\mathcal{M}_{\rm2} + \mathit{i} \mathcal{M}_{\rm3}$, with each
matrix element defined below:
\begin{eqnarray}
i \mathcal{M}_{\rm1} &=& -ie^2 \epsilon_\mu^*(p_3) \epsilon_\nu^*(p_4) \bar
v(p_2) \left[ \frac {\gamma^\nu \not\!q \gamma^\mu}{t} + \frac
{\langle\theta^2\rangle}{96} \nonumber \right. \\ &&\times \left. \frac{t}{2}
\{\gamma^\nu \not\!q \gamma^\mu + p_2^\nu \gamma^\mu - p_1^\mu \gamma^\nu
\} \right] u(p_1),
\end{eqnarray}
\begin{eqnarray}
i \mathcal{M}_{\rm2} &=& -ie^2 \epsilon_\mu^*(p_3) \epsilon_\nu^*(p_4) \bar
v(p_2) \left[ \frac {\gamma^\mu \not\!q_1 \gamma^\nu}{u} + \frac
{\langle\theta^2\rangle}{96} \nonumber \right. \\ && \times \left. \frac{u}{2} \{
\gamma^\mu \not\!q_1 \gamma^\nu + p_2^\mu\gamma^\nu - p_1^\nu \gamma^\mu
\} \right] u(p_1),
\end{eqnarray}
\begin{eqnarray}
i \mathcal{M}_{\rm3} &=& ie^2 \epsilon_\mu^*(p_3) \epsilon_\nu^*(p_4) \frac
{\langle\theta^2\rangle}{192} \bar v(p_2) \nonumber \\ && \times \left[t
\{ p_1^\mu \gamma^\nu - p_2^\nu \gamma^\mu \} + u\{p_1^\nu
\gamma^\mu - p_2^\mu \gamma^\nu \}  \nonumber \right. \\ && + \left.
2(\not\!p_3 - \not\!p_4)(p_1^\nu p_2^\mu - p_1^\mu p_2^\nu) \right]
u(p_1).
\end{eqnarray}\\
It is easy to show that if either one of the polarization vectors is
replaced with its momentum, the matrix element will be zero as we expect
from gauge invariance.  Next it is straightforward to show that the spin
average square matrix element is:
\begin{equation}
\overline{|\mathcal{M}|^{\rm2}} = 2e^4 \left[ \frac{t}{u} + \frac{u}{t}
- \frac{\langle\theta^2\rangle}{96} (t^2 + u^2)\right].
\end{equation}
To first order in $\langle\theta^2\rangle/12$ this gives the following
CM differential cross section:
\begin{equation}
\frac{d\sigma}{d\cos\theta} = \left(
\frac{d\sigma}{d\cos\theta}\right)_{QED} \left[1 - \frac{\langle\theta^2\rangle}{192}
\frac{s^2}{2}\sin^2\theta\right].
\end{equation}

\section{\label{sec:bound}Bounds on $\Lambda_{NC}$ from colliders}

M\o ller scattering experiments do not provide data at high enough energy
to set a bound comparable to the one obtained from Bhabha scattering.  For
Bhabha scattering the bound can be extracted from a series of LEP
experiments \cite{LEPbhabha}.  The total cross section integrated between
$\theta_0$ and $180^\circ-\theta_0$ predicted by our
calculation can be written as:
\begin{equation}
\sigma=\sigma_{SM}+\frac{\pi\alpha^2 s}{8\Lambda^4_{NC}} \{\frac{25}{4}a
+\frac{7}{12}a^3+2\ln\frac{1-a}{1+a}\},
\end{equation}\\
with $a=\cos\theta_0$.  This matches the cut introduced by the L3
experiment where $\theta_0 = 44^\circ$ is the angle relevant to the L3
detector.  Here we use $\sigma_{SM}$ instead of
$\sigma_{QED}$ to take into account the weak interaction and radiative
corrections.  We have neglected the noncommutative correction to
higher order QED and weak interactions.  We use the numerical values of
the data above (TABLE~\ref{tab:table1}) \cite{LEPbhabha}, and for the
theoretical prediction we add the correction due to noncommutativity
obtained in the previous section to the listed SM cross section.  The
$\chi^2$ function is defined as follows:
\begin{equation}
\chi^2 = \sum_i
(\frac{\sigma^i_{exp}-\sigma^i_{theor}}{\Delta^i_{exp}})^2
\end{equation}
with $\Delta^2_{exp}=\Delta^2_{stat}+\Delta^2_{sys}$ and $i$ sums over the
energy range.  Performing the $\chi^2$ analysis over the energy range
shown in TABLE~\ref{tab:table1}, we obtain the bound $\Lambda_{NC}\ge
137$~GeV (95\%C.L.).

\begin{table}
\caption{\label{tab:table1}Bhabha Scattering: Data from L3 experiment at
LEP and SM Prediction.\cite{LEPbhabha}.}
\begin{ruledtabular}
\begin{tabular}{ccc}
 $\sqrt{s}$(GeV) & $\sigma_{exp}\pm\Delta_{stat}\pm\Delta_{sys}$(pb) &
 $\sigma_{SM}$(pb)  \\
\hline
 130.10 & 51.10$\pm$2.90$\pm$0.20 & 56.50 \\
 136.10 & 49.30$\pm$2.90$\pm$0.20 & 50.90 \\
 161.30 & 34.00$\pm$1.90$\pm$1.00 & 35.10 \\
 172.30 & 30.80$\pm$1.90$\pm$0.90 & 30.30 \\
 182.70 & 27.60$\pm$0.70$\pm$0.20 & 26.70 \\
 188.70 & 25.10$\pm$0.40$\pm$0.10 & 24.90 \\
\end{tabular}
\end{ruledtabular}
\end{table}

A similar analysis can be performed on $e^+e^-\to\mu^+\mu^-$ using the
data from the same experiment at LEP \cite{LEPbhabha}.  The total cross
section integrated between $\theta_0$ and $180^\circ-\theta_0$ is:
\begin{equation}
\sigma=\sigma_{SM}+\frac{\pi\alpha^2 s}{8\Lambda^4_{NC}} \frac{a^3}{3},
\end{equation}\\
with $a$ defined above and $\theta_0=44^\circ$.  Fitting our
theoretical prediction to LEP data
(TABLE~\ref{tab:table2}) \cite{LEPbhabha} using $\chi^2$ fit will set
the bound for $\Lambda_{NC}\ge 86$ GeV (95\%C.L.).

\begin{table}
\caption{\label{tab:table2}$e^+e^-\to\mu^+\mu^-$: Data from L3 experiment and
SM Prediction.\cite{LEPbhabha}.}
\begin{ruledtabular}
\begin{tabular}{ccc}
 $\sqrt{s}$(GeV) & $\sigma_{exp}\pm\Delta_{stat}\pm\Delta_{sys}$(pb) & $\sigma_{SM}$(pb) \\
\hline
 130.10 & 21.00$\pm$2.30$\pm$1.00 & 20.90 \\
 136.10 & 17.50$\pm$2.20$\pm$0.90 & 17.80 \\
 161.30 & 12.50$\pm$1.40$\pm$0.50 & 10.90 \\
 172.30 & 9.20$\pm$1.30$\pm$0.40 & 9.20 \\
 182.70 & 7.34$\pm$0.59$\pm$0.27 & 7.90 \\
 188.70 & 7.28$\pm$0.29$\pm$0.19 & 7.29 \\
\end{tabular}
\end{ruledtabular}
\end{table}

For diphoton production, the bound can be extracted from a series of
experiments at LEP\cite{LEPpair}.  The total cross section integrated
between $\theta_0$ and $180^\circ-\theta_0$ predicted by our calculation
can be written as:
\begin{equation}
\sigma=\sigma_{SM}-\frac{\pi\alpha^2 s}{16\Lambda^4_{NC}}
\{a+\frac{a^3}{3}\},
\end{equation}\\
with $a=\cos\theta_0$.  This time the bound is obtained from an analysis
done by the experimenters themselves for the purpose of bounding a generic
contribution for `new physics.'  The bound set from diphoton production
experiments at LEP, as obtained by the DELPHI collaboration and translated
to our definition of noncommutativity scale is $\Lambda_{NC}\ge 160$ GeV
\cite{LEPpair}.  A similar analysis by the L3 collaboration yields a
similar bound \cite{LEPpair}.

A next linear collider (NLC) with a luminosity $3.4\times10^{34}\,{\rm
cm}^{-2}\,{\rm s}^{-1}$ and center of mass energy 1.5~TeV will set a better
bound for $\Lambda_{NC}$.  We calculated the number of events predicted by
ordinary QED at 1.5 TeV and took the statistical uncertainty from the
square root of the number of events.  By requiring the `new physics'
effect to be significant only if it can produce an effect at least 2
standard deviations away from this predicted value, a prediction for the
bound that could be set for the noncommutative scale can be obtained.  Our
calculation for Bhabha scattering predicts a reach for
$\Lambda_{NC}\approx2.0$~TeV, for $e^+e^-\to\mu^+\mu^-$
$\Lambda_{NC}\approx1.7$~TeV, for M\o ller
scattering $\Lambda_{NC}\approx2.7$~TeV and for diphoton production
$\Lambda_{NC}\approx2.0$~TeV.  From this we can conclude that the bound
obtained from these experiments will be about $\approx 2$ TeV and is comparable
to the energy scales where the experiments are performed.

\section{\label{sec:conclusion}Conclusion}
We have considered the phenomenology of a Lorentz-conserving version of
noncommutative QED.  In this theory, spacetime coordinates are promoted to
operators satisfying the DFR Lie algebra.  As opposed to the Lorentz-violating
canonical noncommutative theory, field theory variables have an additional
dependence on the operator $\theta$ which characterizes the
noncommutativity.  This is handled by expanding the fields in powers of
$\theta$, and using gauge invariance and noncommutativity restrictions to 
determine the fields order by order in $\theta$.  Lorentz-invariance
restricts interaction vertices to contain only even powers of $\theta$,
which has distinct consequences on the phenomenology of the theory.  We
considered various $e^{+}e^{-}$ and $e^{-}e^{-}$ collider processes.  The 
cross section was calculated to second order in $\theta$ for Bhabha, M\o
ller, and  $e^{+}e^{-} \to \mu^{+}\mu^{-}$scattering, as well as
$e^{+}e^{-}\to \gamma\gamma$.  Results were then compared to LEP 2 data,
and bounds on the energy scale of noncommutativity, $\Lambda_{NC}$, were
obtained.  The tightest bound came from diphoton production which yielded
$\Lambda_{NC}>$ 160 GeV at the 95\% confidence level.  We also determined
that an NLC running at 1.5 TeV with a luminosity of $3.4 \times 10^{34}\,
{\rm cm}^{-2}\,{\rm s}^{-1}$ will be able to probe $\Lambda_{NC}$ up to
$\sim 2$~TeV.

\begin{acknowledgments}
The authors would like to thank C.D. Carone and C.E. Carlson for inspiring
us to work on this project and also for very helpful discussion during the
project. We thank the National Science Foundation (NSF) for support under 
Grant No. PHY-9900657. In addition, J.M.C. and H.J.K. thank the NSF for 
support under Grant Nos. PHY-0140012 and PHY-0243768.
\end{acknowledgments}

\end {document}